\documentclass[aps,prc,showpacs,amsmath,amssymb]{revtex4}

\usepackage{epsfig}

\begin{document}

%\preprint{BNL-62600}

\author{D.~E.~Kahana}
\affiliation{31 Pembrook Dr.,
Stony Brook, NY 11790, USA}
\author{S.~H.~Kahana}
\affiliation{Physics Department, Brookhaven National Laboratory\\
   Upton, NY 11973, USA}

\title{Elliptical   Flow   in   Relativistic  Ion   Collisions   at
  $\sqrt{s}$= 200 GeV}

\date{\today}  

\begin{abstract}
A consistent picture of the Au+Au and D+Au, $\sqrt{s} = 200$ A GeV
measurements at RHIC obtained with the PHENIX, STAR, PHOBOS and BRAHMS
detectors including both the rapidity and transverse momentum spectra was
previously developed with the simulation LUCIFER. The approach was modeled
on the early production of a fluid of pre-hadrons after the completion of an
initial, phase of high energy interactions. The formation of pre-hadrons is
discussed here, in a perturbative QCD approach as advocated by Kopeliovich,
Nemchik and Schmidt.  In the second phase of LUCIFER, a considerably lower
energy hadron-like cascade ensues.  Since the dominant collisions occurring
in this latter phase are meson-meson in character while the initial
collisions are between baryons, i.~e.  both involve hadron sized interaction
cross-sections, there is good reason to suspect that the observed elliptical
flow will be produced naturally, and this is indeed found to be the case.
\end{abstract}


\maketitle 
\section{Introduction}

RHIC experiments~\cite{phenixflow,phobosflow,starflow} have identified a
transverse momentum asymmetry, designated as `elliptical flow,' and
characterized by the variable $v_2$.  This quantity has deservedly attracted
considerable interest.  Early heavy ion experiments at the LBL
TEVALAC~\cite{lblflow,lbltheory} defined flow in terms of the asymmetry of the
transverse momentum distributions observed for a variety of particles, both
baryons and mesons, produced in relativistic heavy ion collisions at these
relatively low energies. Investigations of flow were subsequently extended to
the AGS heavy ion program~\cite{agsflow,dekflow} and eventually flow was
studied in experiments at the CERN SPS. LBL and even the AGS energies per
nucleon were of course well below those achievable at RHIC.  At lower
energies, however, it was possible to examine an event by event phenomenon,
and to extract what was generally referred to as directed or sidewards
flow~\cite{lblflow}, separately for baryons and mesons. This was of critical
importance in understanding the dynamics of relativistic ion collisions at
those energies.  Much theoretical effort was expended in understanding this
phenomenon in terms of hydrodynamics~\cite{hydro1} employing particular
equations of state for high density nuclear matter and setting some 
set of initial conditions for the hydrodynamics.

Considerable success, however, was also obtained~\cite{dekflow} and a
description of the sidewards flow using only a cascade oriented simulation,
also proved to be possible, with its own inherent treatment of the nuclear
medium.  Here we discuss at $s^{1/2}$ = 200 GeV only the prominent, designated
elliptical, flow extracted from a combined history of many events.  We are
also limited to overall particle distributions, necessarily dominated by final
stable mesons.

Flow may be formally introduced by expressing the observed transverse momentum
distributions as a function of a tranverse azimuthal angle $\phi$, relative to
the angle $\psi_R$ fixing the inclination of initial reaction plane.

\begin{equation}
\frac{dN}{d(\phi-\psi_R)}  =   \frac{1}{2\pi}  \left\lbrack  1  +
\sum_n {2\nu_n \cos (n (\phi - \psi_R))}\right\rbrack
\end{equation}

Of particular interest at RHIC is the second term in this expansion, i.~e.
$v_2$, which if non-vanishing indicates an elliptical asymmetry.  The first
term, if non-zero, would signal the presence of directed or sidewards flow,
and has been estimated in the RHIC detectors
\cite{phenixflow,phobosflow,starflow}, the corresponding parameter $v_1$, is
found to be small in these experiments.  Data has been accumulated by three of
the RHIC detectors~\cite{phenixflow,phobosflow,starflow} producing
experimental studies of the ellipticity for different degrees of centrality,
and also as a function of $p_\perp$.  A comparison is made here between the
PHOBOS measurements and LUCIFER simulations. Reasonable descriptions of both
the observed $DN/D\eta$ spectra in $Au+Au$ as well as the ratios of $p_\perp$
distributions in $D+Au$ and $Au+Au$ at $200$ GeV were already obtained with
LUCIFER simulation.~\cite{luc4brahms,brahms,luciferAuAu}.

We previously cited~\cite{luc4brahms,luciferAuAu} the theoretical difficulties
that arise in producing sufficient flow in a straight-forward parton cascade
models, {\it e.g.} the studies done by Molnar~\cite{molnar}. This author
suggested the possible use of parton coalescence as a mechanism to produce the
otherwise too large observed ellipticity. Otherwise, as Molnar notes, the
parton-parton cross-section must be made as large as $49$ mb~\cite{euroschool}
to yield the measured ellipticities.  In the sense that we are in fact
exploiting the properties of a strongly interacting pre-hadronic fluid at a
second, but still early, stage in our simulation, this hypothesis of
Molnar~\cite{molnar} and our approach do have considerable overlap,
practically speaking.  A relatively large size for the interacting objects
present in the cascade dynamics is surely a key ingredient in reproducing the
required elliptical flow.

We note from the outset, though, that an appreciable degree of flow, perhaps
more than $30\%$, is already generated by the phase I nucleon-nucleon
interactions, and subsequently transmitted to the produced pre-hadrons
colliding in Phase II of the simulation.  These initial interactions are
mediated by the even larger baryon-baryon cross-sections, and the flow in this
initial phase is of course an imprint of the geometrical asymmetry which is
inherent in ion-ion collisions occurring at non-zero impact parameter.

The approach presented here may be of less validity for the lowest $p_\perp$
charged particles, but becomes increasingly accurate with rising $p_\perp$.
Nevertheless we display results from the modeling for even low $p_\perp$.

%One in fact  then expects  a close
%correlation between $v_2(p)$ and the degree of suppression seen
%in the $p_\perp$ spectrum.

In what follows we outline the dynamics behind our earlier calculations which
yielded a successful description of both $D+Au$ and $Au+Au$ data at
$\sqrt{s}=200$ GeV. Of course the same dynamics is employed in the present
determination of the ellipticity.  Subsequently, actual calculations of
elliptical asymmetry are displayed for a variety of kinematic cuts in
centrality and transverse momentum and compared, in particular, to PHOBOS
measurements.

\section{The Simulation}

The simulation code LUCIFER, developed for high energy heavy-ion collisions
has previously been applied to both SPS energies $\sqrt{s}=(17.2,20)$ A
GeV~\cite{lucifer1,lucifer2} and to RHIC energies $\sqrt{s}=(56,130,200)$A GeV
~\cite{luc4brahms,luciferAuAu,luc3}.  Although nominally intended for dealing
with soft, low $p_\perp$, interaction it is possible to introduce high
$p_\perp$ hadron spectra via the NN inputs, which form the building blocks of
the simulation, and to then examine the effect of re-scattering, and
concomitant energy loss, on such spectra~\cite{luc4brahms,luciferAuAu}.  The
simulation is divided into two phases I and II, with most re-scattering and
explicit two body energy loss restricted to the second and considerably
reduced energy stage.  The first stage sets up the participants, both mesonic
and baryonic, their four momenta and positions for the commencement of the
cascade in II, and hence then also involves energy loss through transfer to
the produced particles.  The second stage energy loss and interactions within
a pre-hadronic fluid played a key role in the reproducing the observed
suppression of the $p_\perp$ distribution in our description of $Au+Au$
collisions.

The purpose of describing such high energy collisions without introducing the
parton structure of hadrons, at least for soft processes, is to set a baseline
for judging whether deviations from the simulation measured in experiments
existed and could then signal interesting phenomena.  The dividing line between
soft and hard processes, the latter being in principle described by
perturbative QCD, is not necessarily easy to identify in heavy ion data,
although many authors believe they have accomplished this within a
gluon-saturation picture~\cite{saturation,cgc1,cgc2}.

%This approach has already yielded a consistent description of at
%least the final stage phenomena leading to measurements in both
%$D+Au$~\cite{luc4brahms} and $Au+Au$ at a variety of kinematical
%conditions, energies and transverse momenta. All these results
%follow given only the presence at early times during the ion-ion
%collision of a fluid of `pre-hadrons' having rather generic properties.

\subsection{Pre-hadrons: Production and Hadronisation Time Scales} 

Such a mechanism for the suppression of high transverse momentum jets may seem
to run counter to the conventional pQCD description.  There does exist,
however, some theoretical justification for a picture in which colourless
pre-hadrons, may be produced at rather early times in a AA collision, and then
play a key role in the further development of the system.  First, there is the
work of Shuryak and Zahed~\cite{zahed} on the persistence of hadron-like
states above the critical temperature in a dense quark-gluon medium as well as
similar results from lattice-driven studies on the persistence of the J/$\Psi$
and other special hadronic states~\cite{lattice}.

Recently a much more transparent treatment has been given, on which we now
mainly rely, by Kopeliovich, Nemchik and Schmidt~\cite{boris1} as well as
Berger~\cite{berger}, who directly consider the temporal dynamics of
pre-hadron production from a pQCD point of view.  In particular Kopeliovich
{\it et al.}, outline the fate of leading hadrons from jets, in deep inelastic
scattering (DIS) on massive ions and discuss the relevance to relativistic
heavy ion collisions.  These works have the distinct advantage of relying on
pQCD.  We attempt to reproduce their development as it is germane to our
treatment of NN, NA and AA interactions.

Figure (1), essentially borrowed from Reference(22), displays the production
of a jet, here at least a moderately high energy quark jet, originating in a
pp or pA event, as well as the time scales relevant to the process.
Kopeliovich {\it et al.}  begin by explicitly considering eP or eA, in which
case the initiating particle in the jet creation is an off-shell photon
($\gamma^*$).  In pp or AA one could equally well employ a gluon at the first
vertex.  The kinematics~\cite{boris1} remains essentially the same then for
nuclear induced events and we elaborate this case somewhat.

The essential feature evident in the figure is the production of a colour
neutral pre-hadron after a time $t_p$, from an initially perturbative high
$p_\perp$ quark, by what one could label as coalescence of the leading quark
with an ambient anti-quark.  This coalescence incidentally is most probable
for co-moving anti-quarks, the scarcity of co-movers with increasing $p_\perp$
thus explaining the rapid fall off of even the pp meson spectrum with
increasing $p_\perp$.  Also prominent in this diagram are the early
perturbative gluon radiation.  These radiated gluons may in due course create
further pre-hadrons, and of course there is then concomitant energy loss from
the initial quark.  Further, for leading partons, one expects that the
hadronisation time $t_h$ is considerably longer than $t_p$, involving soft
processes that eventually put the pre-hadron final on the hadronic mass shell.
Octet $q \bar q$'s might also arise but will not persist or eventually
coalesce given the repulsive or non-confining forces in such systems.

One can summarise the kinematic arguments~\cite{boris1} most easily in the
rest frame of the struck proton or nucleus. We use these arguments to estimate
the production time $t_p$ for the pre-hadron, which will be a considerably
shorter time $\tau_p$ in the colliding frame of an AA system. For DIS on a
nucleon, on its own or within a nucleus A, the production time in Figure (1)
is estimated to be

\begin{equation}
t_p \sim \frac{[E_q]}{[dE_q/dz]}(1-z_h),
\end{equation}
\noindent
where $z_h$ is the fraction of quark energy imparted to the hadron.
Integrating the gluon radiation spectrum one obtains for the energy loss per
unit length $z$, a time independent rate~\cite{boris1,niedermayer} rising
quadratically with the hard scale $Q$,

\begin{equation}
-\frac{DE}{Dz} = \frac{2\alpha_s}{3\pi}Q^2.
\end{equation}

The colour neutralisation time is then given by:

\begin{equation}
t_p \sim
\frac{E_q}{Q^2} (1-z_h),
\end{equation}
where $z_h=E_h/E_q$ is the fraction of energy imparted to the hadron.

\noindent
The hadronisation time is related to the QCD scale
factor and is usually estimated as:
\begin{equation}
t_f \sim
\frac{E_h}{\Lambda^2_{QCD}},
\end{equation}

\noindent and is much longer than the colour neutralisation time, given that
the QCD scale is close to 200 GeV.

The authors of Reference (22) argue that the brevity of the de-colourisation
time scale is a consequence of the effects of energy conservation, coherence
and Sudakov suppression.

Importantly, a much reduced time scale for pre-hadron creation is likely to
remain true even for non-leading partons provided Equation (4) remains valid,
{\it i.~e.} when $Q$ is comparable to $E_q$.  Thus many of the radiated hard
or moderately hard gluons, in the early stages of a pp or AA collision will
initiate similar pre-hadrons, and in a nuclear medium such large sized objects
will be numerous and of critical importance to the dynamics.  One should keep
in mind that the overall multiplicity is not large at $p_\perp \ge$ 1 GeV/c,
where the NN $p_\perp$ spectrum, as seen in Figure (2)~\cite{ua1}, has already
dropped well below that at the softest transverse momenta measured: by close
to two orders of magnitude at $p_\perp \sim 1$GeV and for 200 GeV Au+Au as
seen in Figure (3)~\cite{phenixAuAu} by somewhat more.

One concludes that the production and hadronisation processes are to some
degree separate: with generally $t_p$ less than $t_f$ and frequently much
less.  Corresponding time scales in, say, the center of momentum frame for an
A+A system will of course be considerably contracted, given respectively by
$\tau_{p,f}\sim [\Gamma^{-1}]t_{p,f}$.  Although the colourless pre-hadron in
Figure (1) is generated by an initially perturbative process with an
anti-quark, it is the subsequent interactions with other such pre-hadrons that
leads to the observed suppression for mesons of appreciable and even moderate
transverse momentum.  The pre-hadron perturbatively begins life with the
$q\bar q$ at small relative distance, and thus has a small initial size, but
in light of the ambient momenta for such high or even moderately high energy
partons, the transverse diameter rapidly increases to the scale of a typical
hadron. Reference (22) suggests the entire growth to pre-hadronic size occurs
quickly.

One can pursue the evolution of the system of pre-hadrons via a
Glauber-theory~\cite{boris1,glauber} based treatment of their interactions,
or, as we do, via a standard cascade model.  The end result is little
different: what has been labeled jet suppression results.  In Glauber theory
the hadron-sized cross-sections produce strong absorption: hard pre-hadrons
simply do not remain in the final state, they are too often absorbed. In the
cascade described hereafter the pre-hadronic medium is sharply cooled by the
interactions and instead of one hard meson, many softer mesons appear at lower
$p_\perp$.

To obtain the final cascade yields for both D+Au~\cite{luc4brahms} and
Au+Au~\cite{luciferAuAu}, it's essential that the characteristic time $t_p$ is
considerably less than $t_f$.  Indeed, this constraint, and the appropriately
large pre-hadron interaction cross-sections also are critical in generating
the surprisingly large elliptical flow that has been measured at
RHIC~\cite{phenixflow,phobosflow,starflow,molnar}.  Kopeliovich {\it et al.}
provide a perturbative QCD justification for both conditions: early
de-colourisation and large pre-hadron interaction cross-sections.

\subsection{Simulation Dynamics}

For completeness we present a brief overview of the dynamics of our Monte
Carlo simulation, which has in fact already been described extensively in
earlier works~\cite{luciferAuAu,luc4brahms,lucifer2}.  Many other simulations
of heavy ion collisions exist and these are frequently hybrid in nature, using
say string models in the initial
state~\cite{rqmd,rqmd2,bass1,frithjof,capella,werner,ko,ranft} together with
final state hadronic collisions, while some are either purely or partly
partonic~\cite{boal,eskola,wang,wang2,geiger,bass2} in nature.  Our approach
is closest in spirit to that of RQMD and K.~Gallmeister, C.~Greiner, and
Z.~Xu~\cite{greiner} as well as work by W.~Cassing~\cite{cassing} and
G.~Wolschin~\cite{wolschin}.  The latter authors seek to separate initial
perhaps parton dominated processes from hadronic interactions occurring at
some intermediate but not necessarily late time, and/or rely on some hadronic
dynamics.

Whereas in the simulations of the $D+Au$ and $Au+Au$ ion-ion collisions our
purpose was to present a background hadronic setting from which deviations in
observations would signal interesting QCD phenomena, the measured flow, which
is certainly appreciable, seems ab initio to demand a pre-hadronic source. In
stage I in the incoming target and projectile nucleons interactions are
tracked, while in II produced particles, considered as pre-hadrons interact in
a hadron-like cascade.  Basic inputs in the entire
simulation~\cite{lucifer1,lucifer2,luc3} are the measured hadron-hadron
cross-sections~\cite{ua5,ua1,fermilab} as functions of energy and the
production multiplicities taken to conform to KNO~\cite{kno} scaling.  The
elementary NN dynamics is split, as in the classic work of Goulianos on the
phenomenological treatment of NN~\cite{goulianos}, into elastic, single
diffractive (SD) and non-single diffractive NSD).

No energy loss is permitted in I, except at the completion of this stage when
production of pre-hadrons is initiated, the collisions occurring essentially
parallel in time.  A history of each collision is retained, from which one can
construct the particle $p_\perp$ from random walk and in the present case can
also infer, for non-vanishing impact parameter, an asymmetry in the transverse
momentum distributions.  This $p_\perp$ and attached asymmetry can be
appropriately passed on to the pre-hadrons generated at the end of phase I.

The transition from stage I to stage II is accomplished with the guidance of
the two-body inputs. The multiplicities, energies and character of the
produced particles are determined, as well as the initial,
transverse-asymmetric momentum distribution.  The entities produced in I are
principally a set of vector
pre-hadrons, %\rho-like and K^*-$like which interact in stage II
and also after a time $\tau_f$ decay into the normal ``stable'' mesons.  The
decays take account of the vector nature of the pre-hadrons. In this work the
baryons are limited to the usual flavour octet.  the comparison with
elementary NN measurements severely limits any freedom in the mass, number and
spectra of the produced premesons.

Pre-hadrons, which when mesonic presumably consist of a spatially loosely
correlated quark and anti-quark pair, are given a mass spectrum between $m
\sim m_\pi$ and $m \sim 1.2$ GeV, with correspondingly higher upper and lower
limits allowed for pre-hadrons including strange quarks.  The Monte-Carlo
selection of masses is then governed by a Gaussian distribution,

\begin{equation}
P(m)= \exp(-(m-m_0)^2/w^2),
\end{equation}

\noindent with $m_0$ a selected center for the pre-hadron mass distribution
and $w=m_0/4$ the width.  For non-strange mesonic pre-hadrons $m_0 \sim 700$
MeV, and for strange $m_0\sim 950$ MeV. Small changes in $m_0$ and $w$ have
little effect since the code is constrained to fit hadron-hadron data. Too
high an upper limit for $m_0$ would destroy the soft nature expected for most
pre-hadron interactions when they finally decay into `stable' mesons.

To reiterate: the cross-sections in pre-hadronic collisions are taken to be
the same magnitude as hadronic cross-sections, at the same center of mass
energy, thus introducing no additional free parameters into the model. Where
hadronic cross-sections or their energy dependences are inadequately known we
employ straightforward quark counting to determine the cross-section
magnitudes.  Further details on this early stage of the simulation can be
found in earlier work~\cite{luc4brahms,luciferAuAu,lucifer1,lucifer2,luc3}.

It should also be noted that the rather small %\sim 13\%$ rise in
$DN/D\eta$  observed by  PHOBOS~\cite{phobosAuAu130} was  in fact
predicted~\cite{lucifer2}  in our  simulation directly  from the
rise in  multiplicities with energy  inserted into LUCIFER  via the
experimentally constrained  NN inputs. This very  small rise puts
some doubts on the achievement  of a standard QCD plasma at RHIC.
The  increase  of entropy  expected  in  a  phase transition  from
confined to unconfined  partons should show up as  a rather sharp
increase in produced mesons.

\section{Stage II: Final State Cascade}

Stage II is a straightforward two body cascade in which the pre-hadrons, and
of course any normal hadrons present, may interact and decay.  The pre-hadron
decay time, taken at rest or in fact in the colliding Au+Au frame uniformly as
$\tau_f \sim$ 1.0 fm., can be viewed as a hadronisation or formation time.
Appreciable energy having being finally transferred to the produced particles
these `final state' interactions occur at considerably lower energy than the
initial nucleon-nucleon collisions in stage I.  The final cascade of course
conserves energy-momentum and leads to additional transverse energy and meson
production.  For Au+Au, the effect of pre-hadron interactions is appreciable,
greatly increasing multiplicities and total transverse energy, $E_\perp$,
through both production and eventual decay into the stable meson species.
These `final state' interactions of stage II were the principal agent in the
suppression of high $p_\perp$ production~\cite{luc4brahms,luciferAuAu}.

The spatial positions of the particles at this time could be assigned in
various ways.  We have chosen to place the pre-hadrons inside a cylinder,
extending in the transverse and longitudinal directions far enough to contain
the final stage I positions of the initial baryons in the group.  We then
allow the cylinder to evolve freely longitudinally according to the
corresponding momentum distributions, for a fixed time $\tau_p$, {\it i.~e.}
roughly the pre-hadron production time in the colliding frame.  At the end of
$\tau_p$ the total multiplicity of pre-hadrons is limited so that, given
normal hadronic sizes appropriate to meson-meson cross-sections $\sim
(2/3)(4\pi/3) (0.6)^3$ fm$^{3}$, hadrons do not overlap within the cylinder.
Such a limitation in density is consonant with the notion that produced
hadrons can only exist as particles when separated from the interaction region
in which they are generated~\cite{gottfried}.  One may conclude from this that
the pre-hadronic matter acts like, at its creation, as an incompressible
fluid, {\it viz.}  a liquid, as depicted in the earliest calculations with
LUCIFER~\cite{luc4brahms,lucifer1}.

%%The  initial  conditions obtaining  at  the  onset  of the  final
%cascade  could  have,  perhaps,  been  arrived  at  through  more
%traditional  means,  even  through  an initial  partonic  cascade
%followed  by  coalescence.  The  second  stage  would then  still
%proceed as  it does here.  We  can, of course,  ascertain to what
%extent  the elliptical  flow,  arises in  each simulation  stage,
%i.~e.  from initial or from final state interactions.

\section{Flow Determination}

We have run simulations for central, mid-central and peripheral cuts used by
PHOBOS~\cite{phobosflow} to correspond to centralities of 3\%-15\%, 15\%-25\%,
and 25\%-50\% respectively.  Calculated results are displayed in Figure(4)
alongside the PHOBOS determinations.  The theoretical curves shown are best
smooth fits to the LUCIFER results, which of course display statistical
errors.  Given the larger non-statistical experimental errors and the
difficulty of calculation at large $|\eta|$, one probably cannot yet take very
seriously the apparent upturn in $v_2$ at the extreme of the pseudo-rapidity
range.  The agreement between measurement and simulation is quite striking.
With the code already completely specified and constrained by the previous
calculations of rapidity and transverse momenta
spectra~\cite{luc4brahms,luciferAuAu} for both D+Au and Au+Au, there is
nothing to adjust to yield the asymmetric elliptical flow.

Of course we must also confront the dependence of ellipticity on transverse
momentum, $p_\perp$, evidenced in the experimental data. This comparison is
given in Figure(5). The present model, without intervention, handles this
behaviour more than adequately. There is an agreement in both magnitude and
$p_\perp$ variation, in particular the clear saturation of ellipticity above
1-2 GeV/c.  For the moment we cannot follow the functional variation of $v_2$
to very high $p_\perp$, since it requires a considerable increase in
computation time.  Nevertheless, the deviation of our model from a naive
hydrodynamic picture is already evident.  More importantly there is
unambiguous confirmation of the presence of hadron-like degrees of freedom, at
an early stage in the 200 GeV Au+Au collision.  The observed magnitude of flow
could not have obtained without hadron sized interaction cross-sections in
both phase I and II of the simulation.

The non-hydrodynamic saturation of $v_2$ with increasing $p_\perp$ presents an
interesting puzzle.  Our best guess is that at the higher transverse momenta
there is less time for interaction within the stage II hadronic medium and
correspondingly less tying of the ellipticity to the initial geometry for the
highest $p_\perp$ mesons.  Also decays of pre-hadrons play an important role
here: both decreasing the flow at low $p_\perp$ and increasing it at high
$p_\perp$.

%Caveats to  ?????
%these suppositions exist.  First  we require more computing power
%to pursue the asymmetry to much larger $p_\perp$ and secondly the
%first simulation stage  seems to play a somewhat  larger role for
%these higher $p_\perp$'s. (Rewite)

There are two clear lessons to be learned. First that the second phase, the
pre-hadronic cascade, must begin early, at a time $\tau_p \sim
(\Gamma)^{-1}t_p$ with $\Gamma$ the average Lorentz factor for pre-hadrons in
the collision center of velocity frame for the colliding gold nuclei.  This is
in accord with what was found earlier in simulations of the single particle
spectra in d+Au and Au+Au.  Second that large cross-sections must be employed
in phase II and indeed also in the initiating NN collisions in phase I.

The flow is apparently a consequence of and requires large interaction
cross-sections, between objects which cascade starting at early times.  This
picture either obviates or is in accord with the phrase ``strongly interacting
quark-gluon plasma'' which has been popularly employed to described the dense
matter produced in the RHIC energy ion-ion collisions, depending on one's
point of view.  The aptness of the description of the pre-hadron medium as a
liquid was made clear in our earlier work, where the pre-hadron multiplicity
was limited by a constraint that the pre-hadrons could not
overlap~\cite{luc4brahms,luciferAuAu,lucifer2}.

%As           indicated           earlier, the          coalescence
%picture~\cite{molnar,bassmueller}   may   also   give  one   some
%understanding of the rapid  dropoff in the NN $p_\perp$ spectrum,
%which we have  of course in the pre-hadron  application taken as a
%given input to nucleus-nucleus.   One might question the validity
%of using  such a  description for very  high $p_\perp$.  In fact,
%what  one  finds in  NN  is a  rapid  fall  in the  $dN/dp_\perp$
%spectrum, presumably a quantitative statement that coalescence is
%increasingly  unlikely  for  high   $p_\perp$.   This  is  to  be
%expected, coalescence being  realised with higher probability for
%co-movers,  which   become  scarcer  for   increasing  $p_\perp$,
%ocurring  between spatially  correlated comovers.   This strongly
%suggests partonic  coalescence calculations should  themselves be
%applied to  NN, i.~e.  to  explain the observed falling  NN meson
%spectrum.
 
\subsection{Time and Interaction Dependence of Ellipticity.}

By way of a theoretical experiment, we have in Figure (6) explored the
variation in produced flow with the initial delay $\tau_p$ before the start of
phase II and also with the magnitude of the meson-meson cross-section in this
phase II. The results are clear. Halving the cross-sections decidedly reduces
the flow, $v_2$, to a value well below the measured results. A similar effect
occurs when the pre-hadron production time $\tau_p$ is increased.  In a dense
many body system there is a natural screening with the cross-section being
effectively limited by that corresponding to the average distance between
particles.  Should the production of flow occur early on, when the density is
high, the limiting cross-section will be $\sim \pi a^2$ with $a$ the average
interparticle distance.  As a result little effect will be evident from a
cross-section increase.

The $\tau_p$ dependence exhibited in Figure(6) is striking, the flow
diminishes very appreciably with a large delay in the commencement of phase
II.  The geometrical, free expansion of the system   reduces
both the densities and the interaction frequencies, thus increasing
the mean free path for pre-hadrons; in hydrodynamic terms this lessens the
pressure which produces the flow.  The connection between ellipticity and the
initial spatial configuration of the ion-ion collision is thus lost by the
time the interactions begin.  One should note that the cascade calculations in
an appropriate limit should yield hydrodynamics.  Of course the hydronamic
limit for flow is achieved experimentally only for low $p_\perp$ and in
Figure(5) both measurements and LUCIFER simulation drop well below the
earliest hydrodynamic predictions at higher low $p_\perp\sim$ 2 GeV/c.

Very soft early gluons and quarks are not a part of the colour neutral medium,
but their eventual hadronisation is approximated by our treatment of soft
hadronic processes, guided by known hadron-hadron interactions and production
multiplicities.  Finally, completely eliminating phase II reduces the
ellipticity $v_2$ by some 25-30\%, depending on the centrality, the remaining
flow being ascribable to the initial nucleon-nucleon interaction which already
occurred in phase I.

\section{Comments}

Overall the very natural fashion in which the elliptical flow is produced in
the model speaks strongly to the important role played by pre-hadrons, {\it
  i.~e.}  colour neutral objects having hadron-sized interaction
cross-sections, and their presence from an early time in Au+Au collisions at
RHIC energies. Elliptical flow is, after all, the one truly collective
phenomenon to have been observed at RHIC, and it is precisely such collective
properties that may indicate the existence of some medium or fluid, in this
case a plasma apparently characterized by objects having hadron sized
interaction cross-sections during much of its existence.  The very earliest
times in an A+A collision, tracked in the first stage of our simulation, also
play a non-negligible role in the production of elliptical flow.  The
nucleon-nucleon interactions occurring in this stage likely lead to a
Cronin-like~\cite{cronin} enhancement~\cite{luciferAuAu} of the high $p_\perp$
spectrum and also transmit the spatial geometry of the collision to the
produced pre-hadrons.  It would be interesting to evolve this first stage with
allowances made for mutual crossing between parallel quark-gluon and hadronic
simulations.  Future work will pursue such an avenue. Also, in agremeent with
the PHOBOS observations~\cite{phobosflow}, the simulation produces only very
small direct flow, i.~e. $v_1$, at $s^{1/2}= 200$ GeV.

\section{Acknowledgements}
This  manuscript  has  been  authored  under  the  US  DOE  grant
NO. DE-AC02-98CH10886. One of  the authors (SHK) is also grateful
to  the  Alexander von  Humboldt  Foundation,  Bonn, Germany  for
continued  support and hospitality.   Useful discussion  with the
BRAHMS,  PHENIX, PHOBOS  and STAR  collaborations  are gratefully
acknowledged, especially  with C.~Chasman, R.~Debbe, F.~Videbaek.
D.~Morrison, M.~T.~Tannenbaum, T.~Ulrich and J.~Dunlop.

\vfill\eject

\begin{figure}
\vbox{\hbox to\hsize{\hfil
\epsfxsize=6.1truein\epsffile[0 0 561 751]{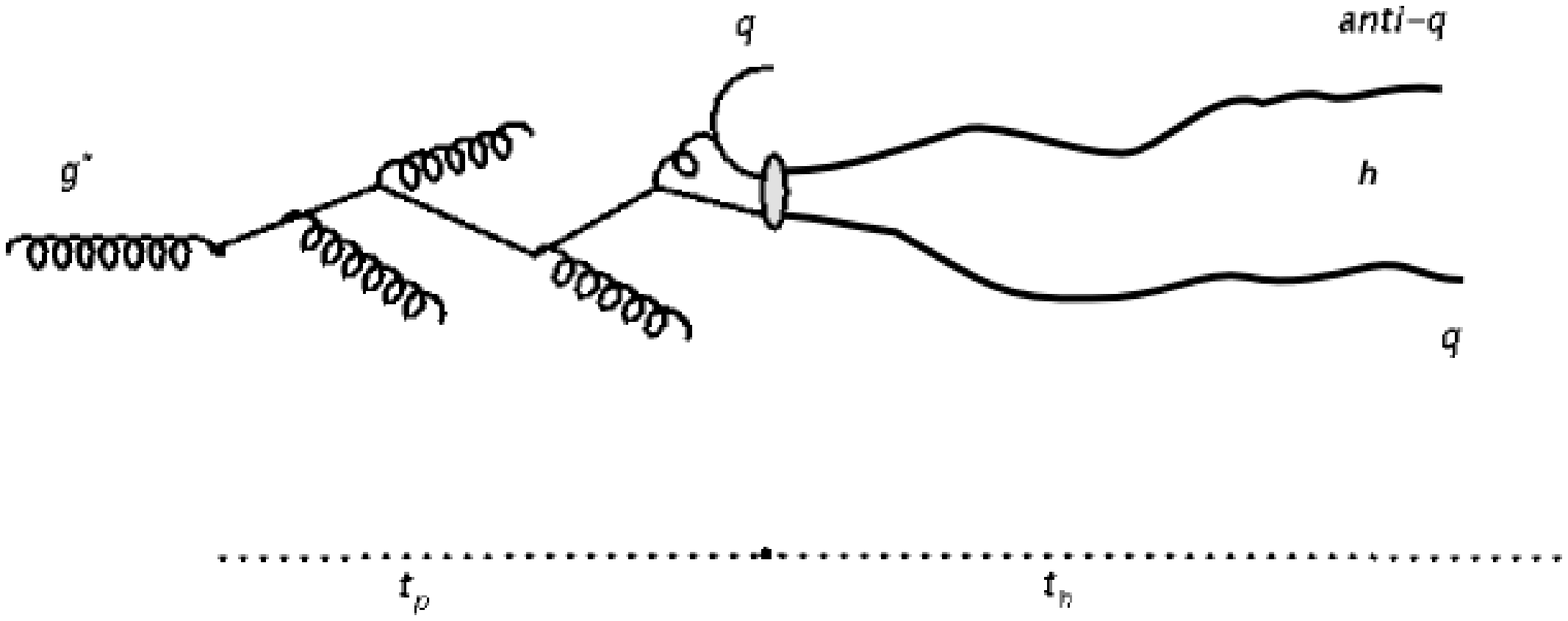}
\hfil}}
\caption[]{Schematic drawing, borrowed from Kopeliovich et
al. ~\cite{boris1}, of the perturbative formation of a pre-hadron from
an off-shell gluon $g^*$ incident on a quark in the rest frame of
nucleus B in an A+B or P+B event.  The quark radiates gluons and
eventually combines with a perturbatively produced anti-quark to form
at first a small colourless pre-hadron which rapidly expands to
hadronic size.  The time $t_p$ signifies the production time of the
pre-hadron and $t_h$ its later hadronisation time. It is argued in the
text that in general $t_p$ is much less than $t_h$, and the existence
of two such time scales is critical to the parton--hadron dynamics.}
\label{fig:Fig.(1)}
\end{figure}
\clearpage

\begin{figure}
\vbox{\hbox to\hsize{\hfil
\epsfxsize=6.1truein\epsffile[0 0 561 751]{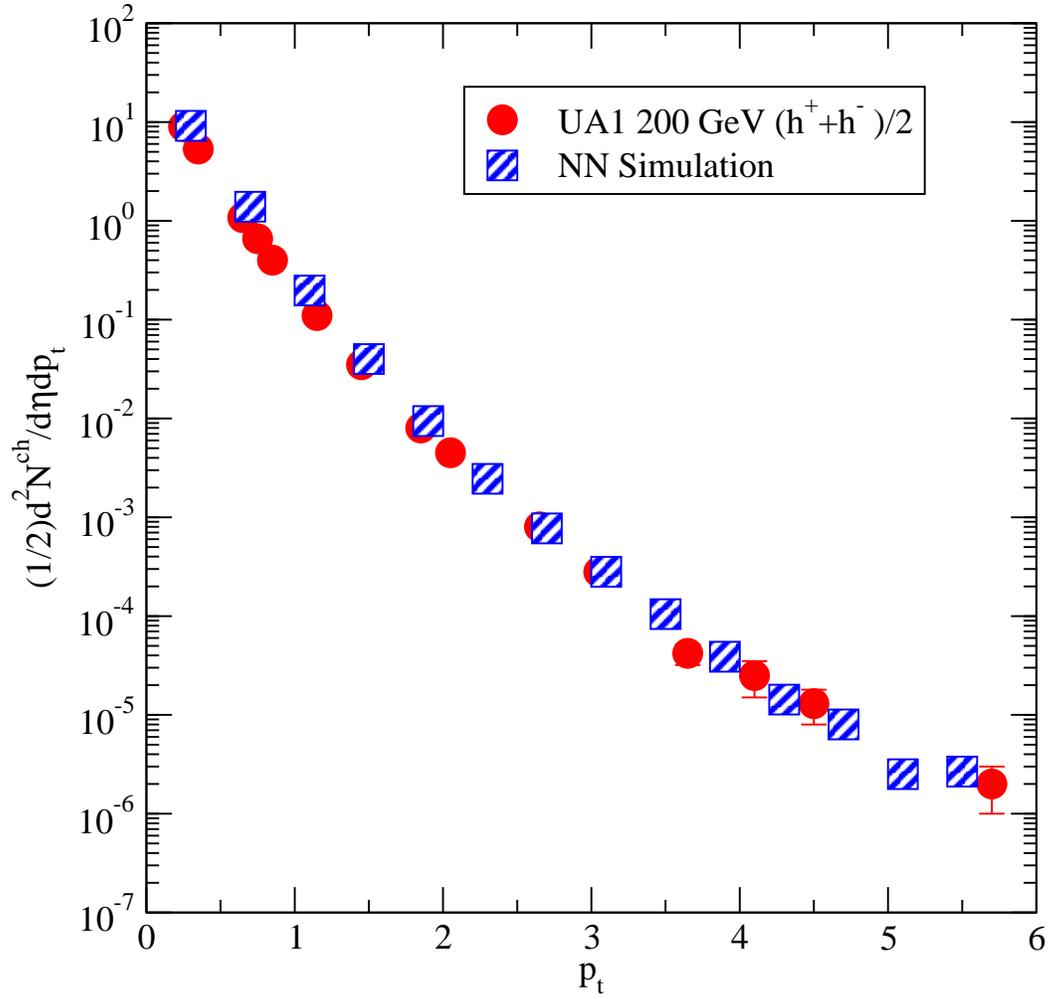}
\hfil}}
\caption[]{pp Pseudo-rapidity spectra: Comparison of UA1 minimum bias
200 GeV NSD data~\cite{ua1} with an appropriate LUCIFER
simulation. The latter is properly constrained by experiment and is an
input to the ensuing AA collisions; thus does not constitute a `set'
of free parameters.}
\label{fig:Fig.(2)}
\end{figure}
\clearpage

\begin{figure}
\vbox{\hbox to\hsize{\hfil  
\epsfxsize=6.1truein\epsffile[0 0 561 751]{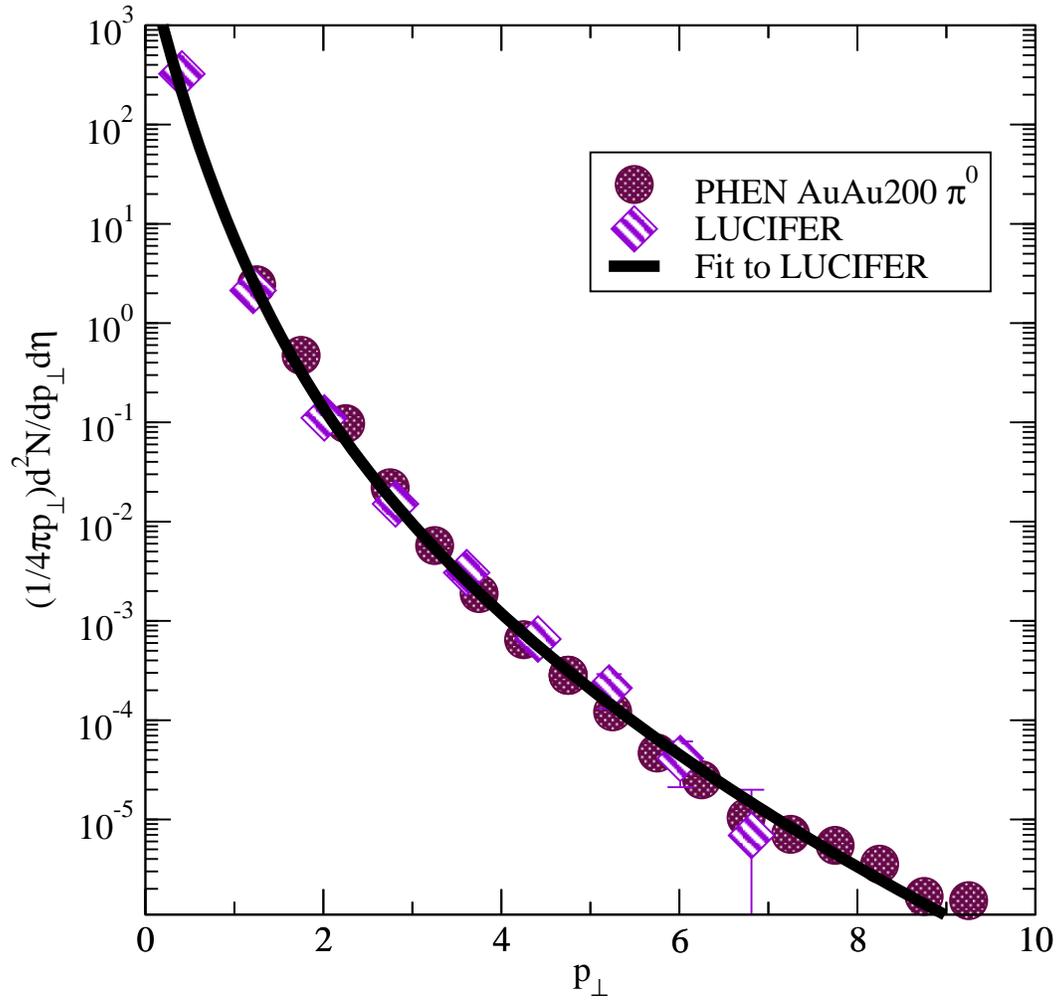} 
\hfil}}
\caption[]{Central PHENIX $\pi^0$ 200 GeV for Au+Au vs simulation.
Curves for different choices of the production time $\tau_p$ differ
very little, since in effect the cascade effectively begins somewhat
later, near $0.25-0.35$ fm/c and continues much longer to tens of
fm/c.  Centrality for PHENIX is here $0\%-10\%$, roughly for impact
parameters $b<4.25$ fm. in the simulation.}
\label{fig:Fig.(3)}
\end{figure}
\clearpage

\begin{figure}
\vbox{\hbox to\hsize{\hfil  
\epsfxsize=6.1truein\epsffile[0 0 561 751]{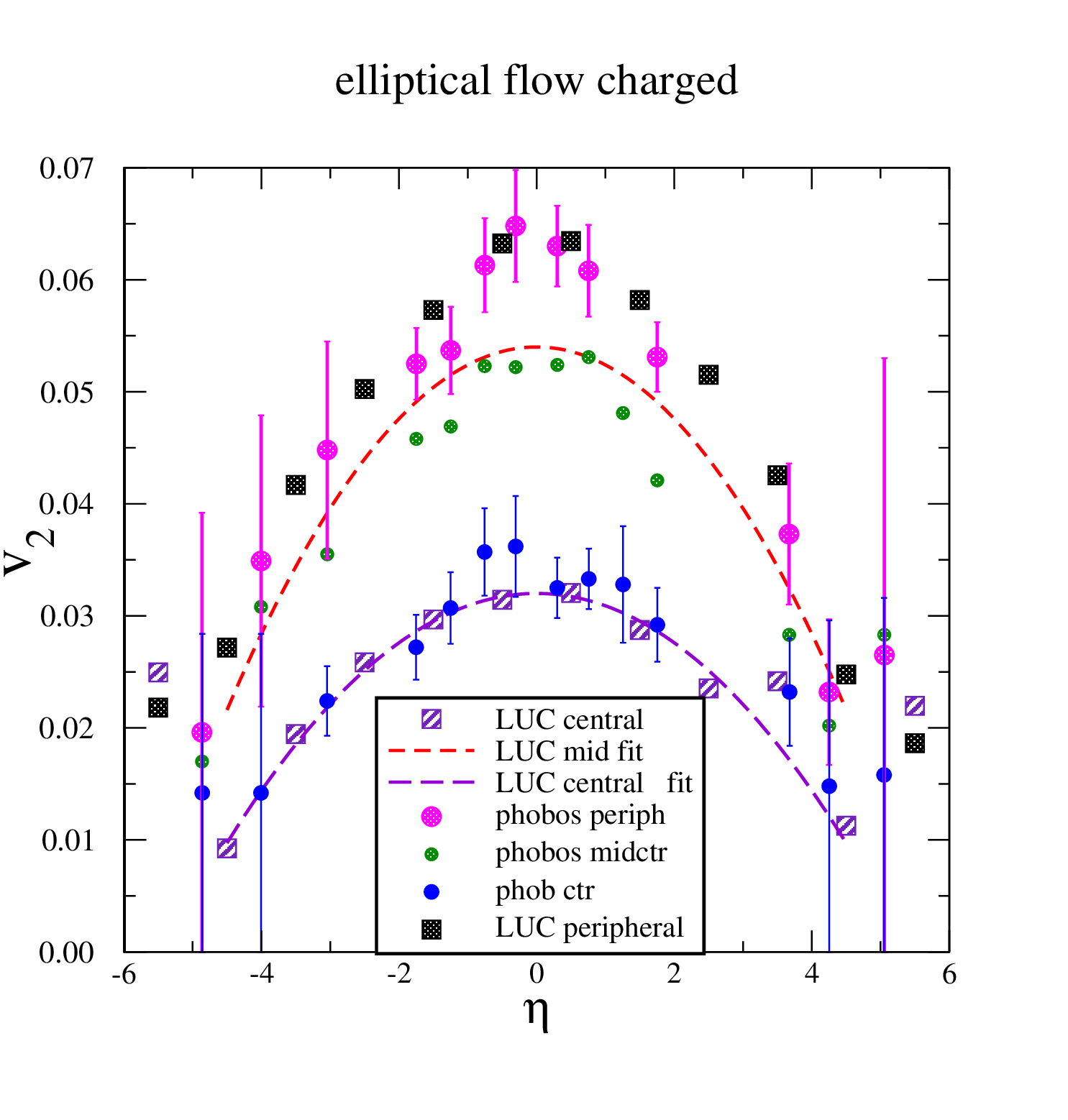} 
\hfil}}
\caption[]{PHOBOS 200 GeV charged flow, $v_2$, as a function of pseudo-rapidity
  for peripheral, mid-central, and central  versus
  LUCIFER simulations, with actual points calculated indicated in central
  and peripheral but for an already busy graph only a fit is shown
  for mid-central.}
\label{fig:Fig.(4)}
\end{figure}
\clearpage

\begin{figure}
\vbox{\hbox to\hsize{\hfil  
\epsfxsize=6.1truein\epsffile[0 0 561 751]{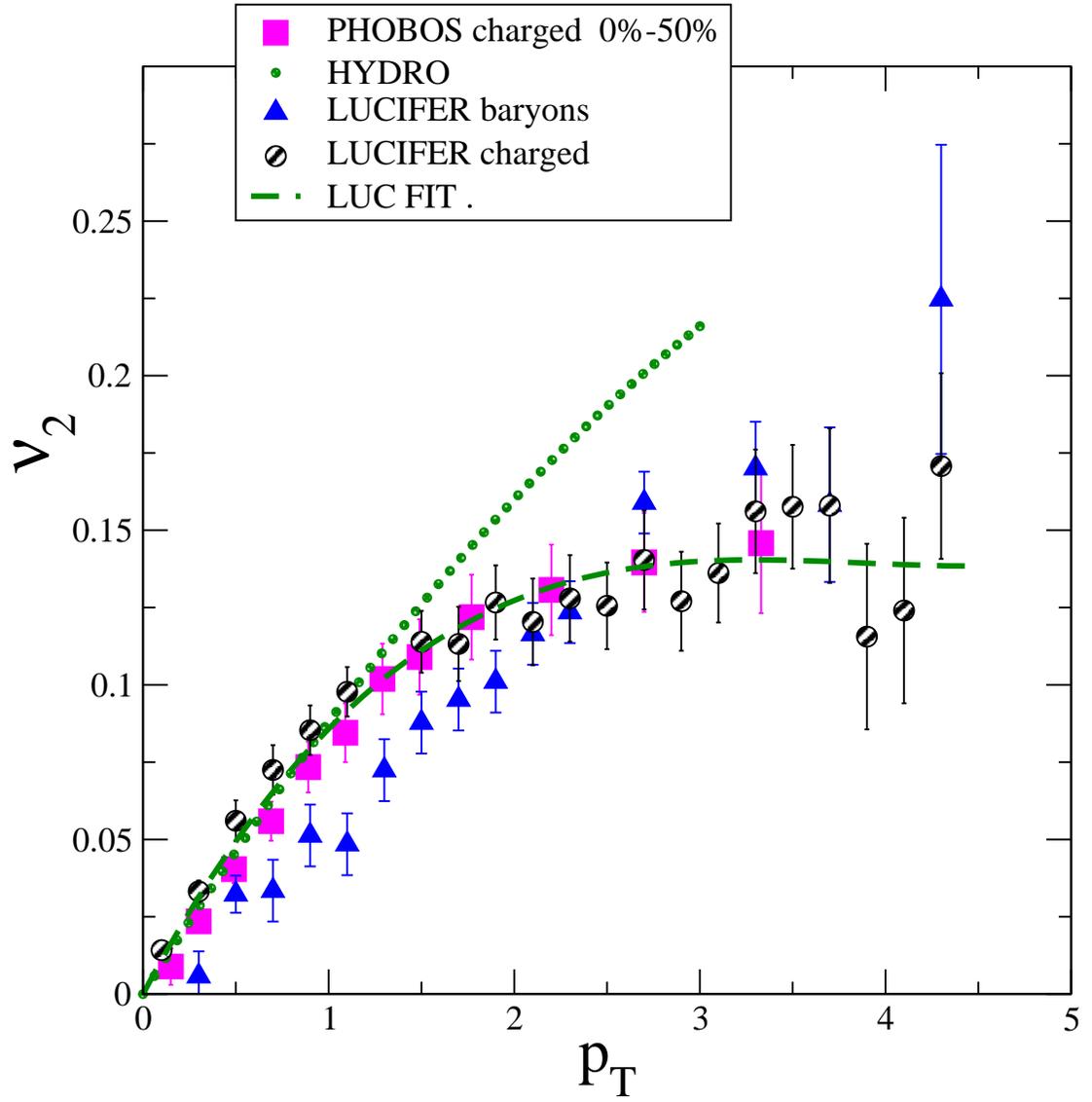} 
\hfil}}
\caption[]{PHOBOS 200 GeV $v_2$ as a function of transverse momentum
  for 0--50\% centrality versus
  LUCIFER simulations for both charged mesons and baryons. Hydro
  calculations are also indicated~\cite{hydro1}.}
\label{fig:Fig.(5)} 
\end{figure}
\clearpage

\begin{figure}
\vbox{\hbox to\hsize{\hfil  
\epsfxsize=6.1truein\epsffile[0 0 561 751]{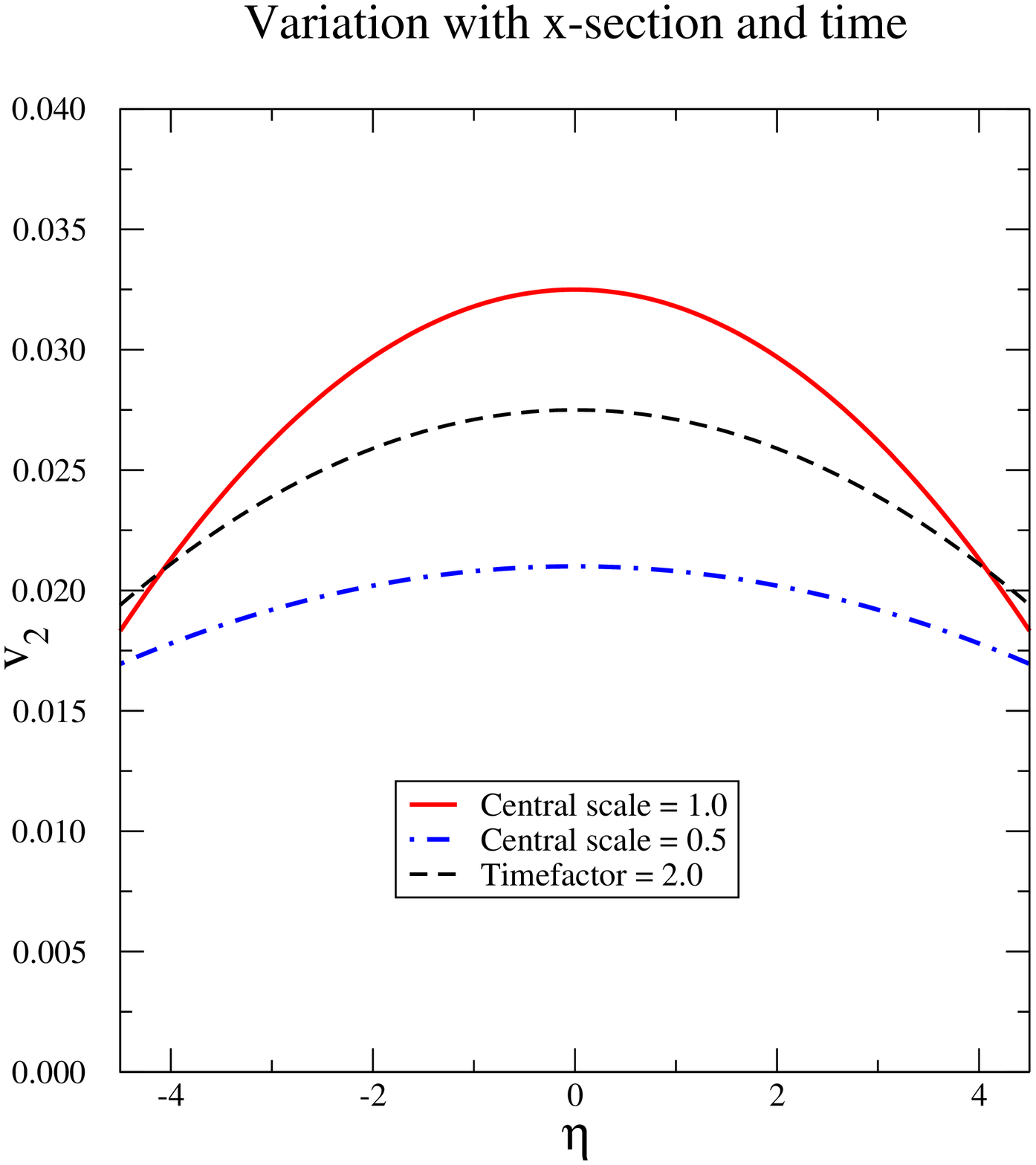} 
\hfil}}
\caption[]{Variation in LUCIFER central simulations with 
  scale changes for the overall pre-hadron-pre-hadron cross section
  by a factor of one-half and the time factor by 2.0. This
  decrease in interaction strength, especially, and lengthening of the time
  lead to considerable reduction in the elliptical flow.}
\label{fig:Fig.(6)}
\end{figure}
\clearpage

\end{document}